\documentclass[11pt]{article}\include{ddefs}
\usepackage{graphics,pifont,array}
\usepackage{graphicx}
\begin{document}

\begin{titlepage}

\begin{center}

{\Large{        \mbox{   }                          \\
                \mbox{   }                          \\
                \mbox{   }                          \\
                \mbox{   }                          \\
                \mbox{   }                          \\
                \mbox{   }                          \\
                \mbox{   }                          \\
                \mbox{   }                          \\
                \mbox{   }                          \\
                \mbox{   }                          \\
  {\textbf{WEDGE DIFFRACTION AS AN INSTANCE OF      \\
          RADIATIVE SHIELDING}}                     \\
               \mbox{    }                          \\
               \mbox{    }                          \\
              J. A. Grzesik                         \\
           Allwave Corporation                      \\
        3860 Del Amo Boulevard                      \\
                Suite 404                           \\
           Torrance, CA 90503                       \\
                \mbox{    }                         \\
           (818) 749-3602                           \\ 
            jan.grzesik@hotmail.com                 \\
         jan.alex.grzesik@gmail.com                 \\
              \mbox{     }                          \\
              \mbox{     }                          \\  
              \today                                      }  }

\end{center}

\end{titlepage}

\setcounter{page}{2}

\pagenumbering{roman}
\setcounter{page}{2}
\vspace*{+1.825in}


\begin{abstract}
\parindent=0.245in
    The celebrated Sommerfeld wedge diffraction solution is re\"{e}xamined from a null
interior field perspective.  Exact surface currents provided by that solution, when
considered as disembodied half-plane laminae radiating into an ambient, uniform space
both inside and outside the wedge proper, do succeed in reconstituting both a specular,
mirror field above the exposed face, and a shielding plane-wave field of a sign opposite
to that of the incoming excitation which, under superposition, creates both the classical,
geometric-optics shadow, and a strictly null interior field at the dominant, plane-wave
level.  Both mirror and shadow radiated fields are controlled by the residue at just one
simple pole encountered during a spectral radiative field assembly, fixed in place by
incidence direction $\phi_{0}$ as measured from the exposed
face.  The radiated fields further provide diffractive contributions drawn from two
saddle points that track observation angle $\phi.$  Even these, more or less asymptotic
contributions, are found to cancel exactly within the wedge interior, while, on the
outside, they recover in its every detail the canonical structure lying at the base of
GTD (geometric theory of diffraction).  It is earnestly hoped that this revised scattering
viewpoint, while leaving intact all details of the existing solution, will impart to it
a fresh, physically robust meaning.  Moreover, inasmuch as this viewpoint confirms,
admittedly in an extreme limit, the concept of field self-consistency (known in rather more
picturesque language as Ewald-Oseen extinction), perhaps such explicit vindication may
yet encourage efforts to seek exact solutions to scattering/diffraction by
electromagnetically permeable (i.e., dielectric) wedges, efforts that harness integral
equations with polarization/ohmic currents distributed throughout wedge volumes as
sources radiating into an ambient, uniform reference medium.        
\end{abstract}


\pagestyle{plain}

\parindent=0.5in

\newpage

\pagenumbering{arabic}

\pagestyle{myheadings}

\setlength{\parindent}{0pt}

\pagestyle{plain}

\parindent=0.5in

\newpage
\mbox{   }

\pagestyle{myheadings}

\markright{J. A. Grzesik \\ wedge diffraction as an instance of radiative shielding}

\section{Introduction}
\vspace{-2mm}

     This paper seeks to add one small nugget of physical interpretation to the celebrated
     Sommerfeld solution of electromagnetic diffraction by a perfectly conducting wedge.
     The elegance of Sommerfeld's symmetry/contour integral solution has proved to be an
     inextinguishable lure to droves upon droves of other researchers who have complemented
     it with a multitude of successful attacks along still other lines of comparable ingenuity.
     The material below makes no pretense whatsoever to extending this theoretical
     repertoire.  On the contrary, it accepts the outcome of the existing Sommerfeld
     theory with the object of imparting to it a modicum of physical insight patently
     deficient in the overabundant wedge diffraction literature.  An informal guide,
     doubtless incomplete, to wedge diffraction theory appears at paper's end.
     
     What we have in mind is to accept the electric surface currents $K$ which the
     Sommerfeld formalism requires to flow upon both exposed and, generally speaking,
     shadow faces, and to build up the fields, both near and far, radiated by these
     currents.  The total electromagnetic field existing everywhere is then regarded
     as having these radiated fields superposed upon the primary, invariably plane
     wave excitation.  The physical feature which may, regrettably, provoke an initial
     urge to repudiate,\footnote{Recoil of this sort is fashioned largely on the basis
     of mechanical experience, which insists that solid bodies resist
     penetration.  The weight of this experience, routinely reinforced throughout
     our lives, automatically spills over into electrodynamics, wherein it initially
     cries out for a similar interdiction.
     Such anxieties should be lessened by recognizing that, unlike its mechanical
     counterpart, electrodynamics ignores hardness {\em{per se}} and responds only
     to charge/current sources, idealized
     herein as being confined to infinitely thin surface sheets.} is that all such
     fields are regarded as being fully able to penetrate into the wedge metallic
     interior wherein, of course, it is only their sum which is required to vanish
     identically.  More than that, the radiated fields are obliged to produce not
     only the specularly reflected wave, but to annihilate also the incoming field
     throughout the traditional shadow region while maintaining therein nonvanishing
     diffractive contributions.  We propose in the sequel to demonstrate all such
     features with unambiguous, albeit not entirely trivial analysis.
     
     The null interior field viewpoint which we advocate here has its precedent in,
     and is indeed but an extreme limit of, the traditional Ewald-Oseen extinction
     phenomenon, wherein a field impinging upon an electrically permeable obstacle
     stimulates throughout its interior a polarization/ohmic current distribution
     which then radiates in its turn in such a manner as to replace (read:
     extinguish) the incoming, ambient-space field with one more suited to propagation
     within the (not necessarily uniform) dielectric material.  When perfectly conducting
     metallic obstacles, such as our present wedge, enter into play, the Ewald-Oseen
     apparatus simply defaults to infinitely thin surface current veneers.

        One finds in [{\bf{1}}] a collection of rather standard electrodynamic problems
     viewed under the null interior field aspect.  That collection begins indeed on a
     humble, electrostatic note having a point charge placed in proximity to a perfectly
     conducting half space, a situation instantly resolved through appeal to the construct
     of a fictitious interior image of equal magnitude and opposite sign.  That formal
     solution, its gentle ingenuity and ready success being given their deserved due,
     provides an access to a real charge distribution across the half space boundary,
     a distribution on whose basis one calculates a potential reproducing that of the said
     image on the exterior, and uniformly cancelling the primary field within.  So fortified,
     and with additional plane/cylinder successes in a genuinely electrodynamic setting,
     a still further problem on loop excitation of a circular waveguide is quick to show
     that the null interior field viewpoint (in this latter case,
     \newpage
     \mbox{   }
     \mbox{   }
     \newline
     \newline
     \newline
     being actually extended
     across the waveguide metallic exterior) can also be harnessed as an active path to
     problem solution, and not merely in the r{\^{o}}le, as here, of an {\em{ex
     post facto}} field dissection.  In fact, the loop current evolution in time is
     taken there to be quite general, its details subsumed under a Laplace transform.
     \vspace{-3mm}

          The analysis now given had its genesis in just such a burgeoning
     Ewald-Oseen background, an analysis which, admittedly, represents the second tier
     of retrenchment from more lofty ambitions.  Initial hopes for success in the use of
     the Ewald-Oseen principle as an active solution basis for the wedge soon foundered,
     first when deployed against the permeable wedge, and then against its present,
     metallic limit.  Efforts to salvage at least some of this work have thus nucleated
     into the present, far more modest {\em{post mortem}} interpretation.  Nevertheless
     it is our hope that even this, vastly reduced program may still enjoy some minor
     scientific merit.  It had been presented quite some time ago, under an identical
     title, at the 1999 IEEE Antennas and Propagation/URSI Symposium in Orlando, Florida
     [{\bf{2}}], but had to await an epoch of sustained leisure to unshackle it from a
     necessarily {\em{d\'{e}p\^{e}che}} format.
     \vspace{-6mm}
     
     \section{Sommerfeld solution summarized}
     \vspace{-5mm}
     
         Sommerfeld's initial diffractive solution for a half-plane, a degenerate wedge
     having $2\pi$ as its exterior angle, appeared in [{\bf{3}}].  In [{\bf{4}}], H. M.
     MacDonald bypassed Sommerfeld's contour integral {\em{cum}} symmetry arguments in
     favor of a development in eigenfunctions {\em{a priori}} adapted to the required
     boundary conditions.  He was thus able to cope with wedge diffraction at arbitrary
     exterior angles short of $2\pi,$ and then Sommerfeld yet again in [{\bf{5}}] provided
     a mature overview of his evolved theory.  Widely available also is Sommerfeld's
     half-plane solution in his volume on optics [{\bf{6}}].  Unburdened by the German
     language barrier of [{\bf{3}}] and [{\bf{5}}], [{\bf{6}}] is available in English
     translation.
     
         The wedge diffraction geometry is given in Figure 1, with exterior angle $\gamma$
     (reckoned from the horizontal face)    
     and a plane wave excitation $E_{z,inc}(r,\phi)=+\exp(-ikr\cos\{\phi-\phi_{0}\})$
     assumed to be incident from above at azimuth $\phi_{0}.$
     Guided by common intuition, we expect the primary excitation to vanish throughout shadow
     region $III,$ to persist intact across region $\rule{2mm}{0mm}I\,\cup\,II,\rule{2mm}{0mm}$
     and to be accompanied across region $I$ by a specularly reflected, mirror companion
     $\rule{1mm}{0mm}E_{z,mirr}(r,\phi)=-\exp(-ikr\cos\{\phi+\phi_{0}\}).\rule{2mm}{0mm}$  Diffractive
     \vspace{-0.35in}
     \begin{center}
     	\includegraphics[angle=-86.5,width=0.40\linewidth]{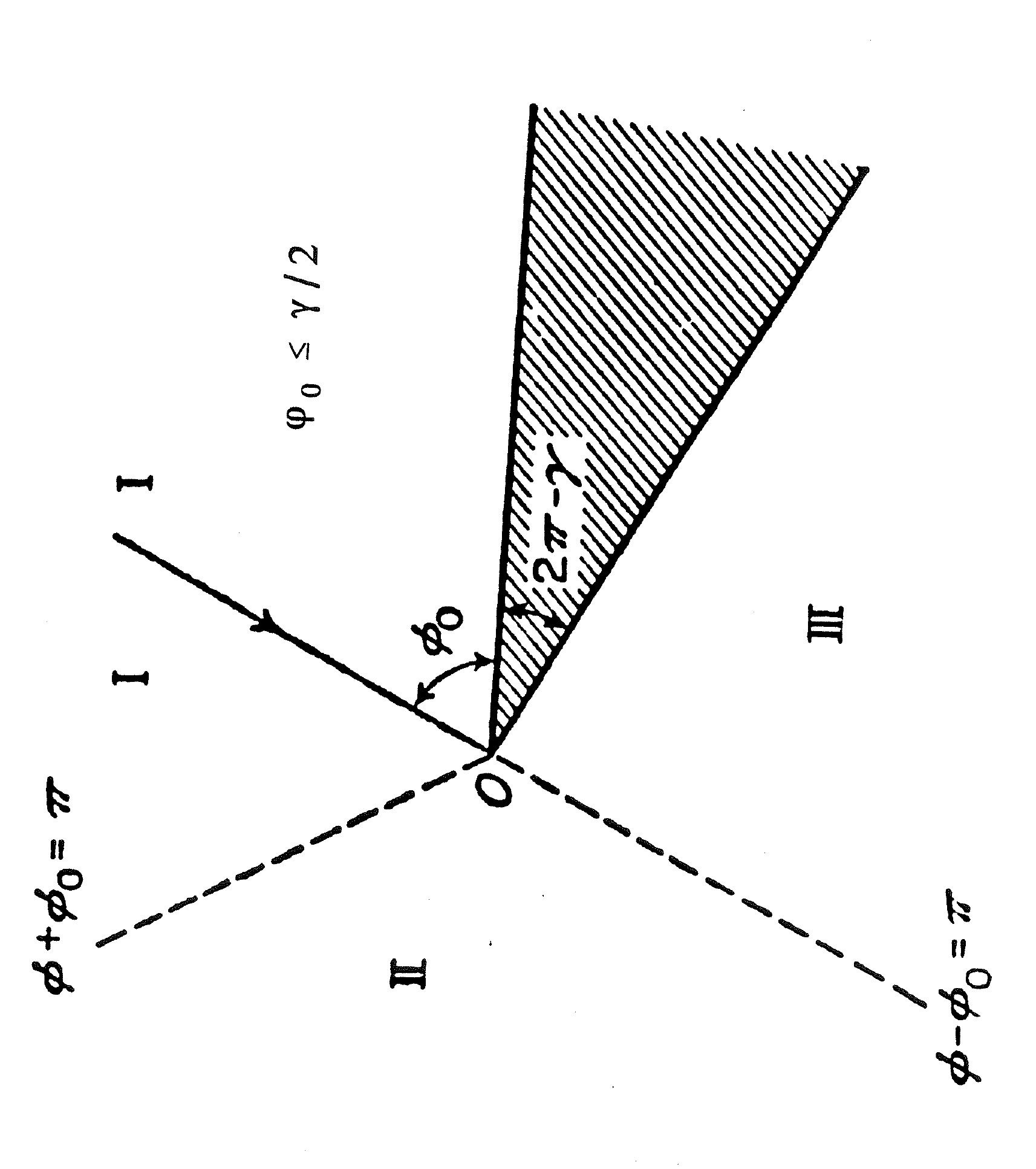}
     \end{center}
     \vspace{-8mm}
     \begin{center}
     	Figure 1.  Diffracting wedge geometry.
     \end{center}        
     \newpage
     \mbox{   }
     \mbox{   }
     \newline
     \newline
     \newline
     \parindent=0in
     contributions are expected to creep into all three regions $I,$ $II,$ and $III,$ whereas the
     perfectly conducting wedge 
     material occupying angular range $\gamma<\phi< 2\pi$ is to be devoid of any electromagnetic
     penetration.  In this simplest of all proof-of-principle problem incarnations, we posit that the
     incidence direction is perpendicular to the edge, which latter naturally serves as the origin $r=0$
     of radial co\"{o}rdinate $r$ complementary to azimuth $\phi,$ while $z,$ measured positive toward
     the reader, completes the co\"{o}rdinate triad.  Of the two canonical field polarizations,
     either electric $E$ or magnetic $H$ parallel to the edge, we limit ourselves only to the former.
     \parindent=0.5in
     
          As an {\em{ansatz}} of great power Sommerfeld introduced the function\footnote{In standard notation,
      $k=\omega/c,$ with $c$ being the speed of light and $\omega$ the angular frequency, taken positive in
      conformity with integral convergence in the shaded half-strips in Figure 2.  Factor $e^{-i\omega t},$
      which governs the field temporal variation across the board, is implicitly acknowledged but otherwise
      hidden from view.}      	
     \begin{equation}
     v(r,\psi)=\frac{1}{2\gamma}\int_{\atop{\rule{-4mm}{5mm}C_{1}\cup\,C_{2}}}
     e^{-ikr\cos\zeta}\,\frac{d\zeta}{1-e^{-i\pi(\zeta+\psi)/\gamma}}     
     \end{equation}       
     {\em{vis-\`{a}-vis}} the $\zeta$-plane contours shown in Figure 2.  In terms of $v(r,\psi)$ the
     field pattern $u(r,\phi),$ both near
     \vspace{-6mm}      
     \begin{center}
     	\includegraphics[angle=0.52,width=0.40\linewidth]{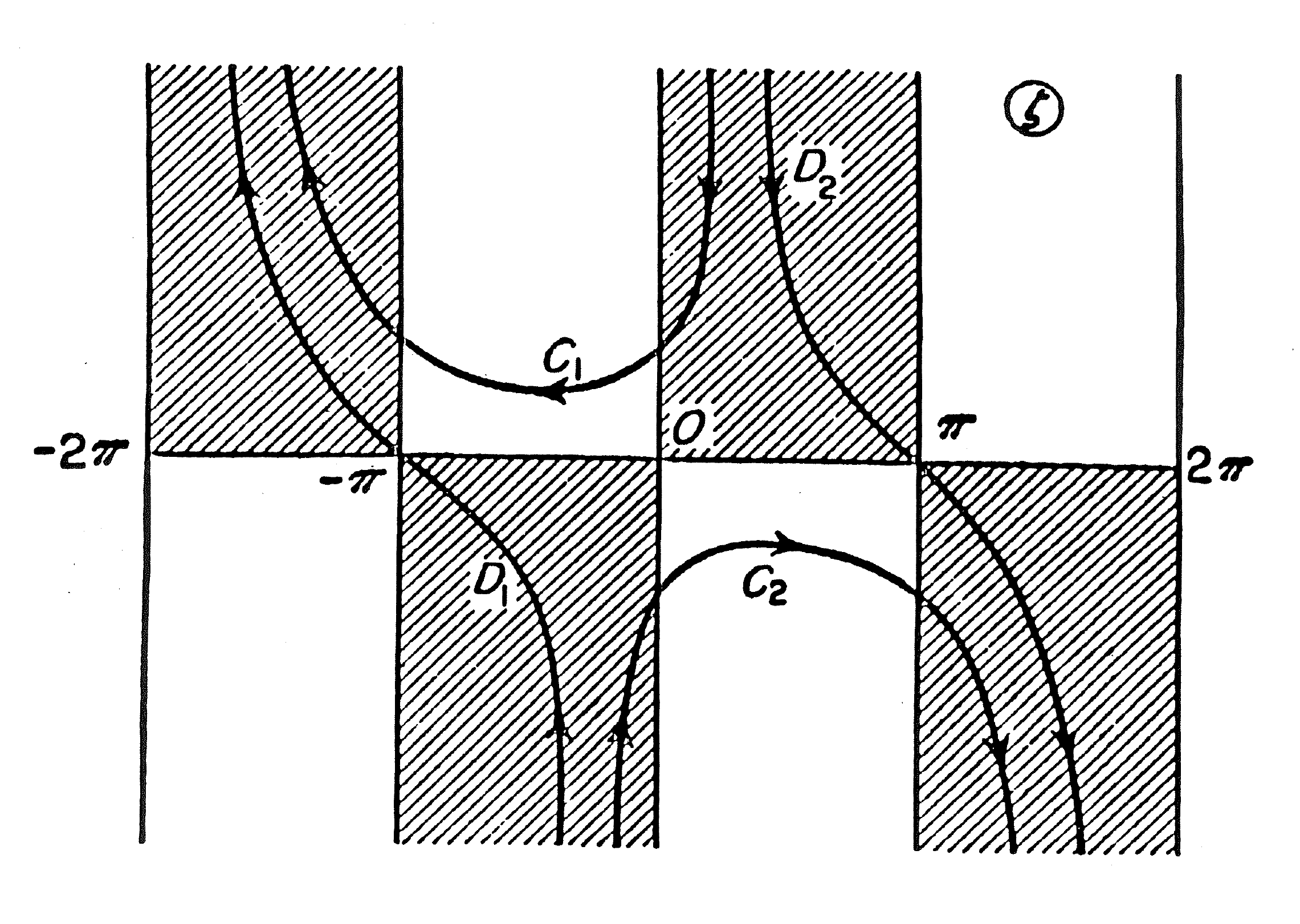}
     \end{center}                   
     \vspace{-11mm}
     \begin{center}
     	Figure 2.  Sommerfeld canonical contours.
     \end{center}
     \vspace{-4mm}
     and far, respectively for the $E/H$ polarizations is obtained as
     \begin{equation}
     u(r,\phi)=v(r,\phi-\phi_{0})\mp v(r,\phi+\phi_{0}) \,.
     \end{equation}
     Our concern henceforth will be exclusively with the upper, minus sign in (2), which provides for
     a null electric field on both wedge faces at $\phi=0$ and $\phi=\gamma.$  It is to be emphasized
     that (2) represents the entire field, incident plus that radiated by surface currents $K$ induced
     on wedge faces.
     
          Figure 1 proclaims the self-evident angular symmetry whereby nothing really new can be
     encountered by allowing incidence angle $\phi_{0}$ to exceed $\gamma/2.$  But even before that,
     with $\gamma-\pi<\phi_{0}<\gamma/2,$ it is possible to have both upper and lower wedge faces
     exposed to the primary illumination, a physically acceptable scenario, to be sure, but one
     that would complicate the ensuing discussion.  Hence, simply on grounds of convenience, we
     legislate that in fact $0<\phi_{0}<\gamma-\pi.$
     
          Contour $C_{1}\cup\,C_{2}$ can of course be deformed into $D_{1}\cup D_{2}$ provided that
     one accounts for the residues at any poles that may crop up within the interval $-\pi<\Re\,\zeta<\pi,$
     residues whose \mbox{form is that of plane waves}
     \newpage
     \mbox{   }
     \mbox{   }
     \mbox{   }
     \newline
     \newline
     representing both incoming and all possible specularly
     reflected contributions.  In this way, on the basis of the paragraph preceding, we wish to
     contend with at most one pole in connection with each of $v(r,\phi\mp\phi_{0}),$ something
     now assured by the demand that $0<\phi_{0}<\gamma-\pi.$  But, while these precautions do
     retain an {\em{a priori}} validity in the context of contour
     $\rule{0mm}{0mm}C_{1}\cup C_{2}\rule{0mm}{0mm}$, they
     are destined to be shortly superseded by considerations which attach to the alternate
     contours illustrated in Figures 3-5.
     \vspace{-2mm}
     
          Function $u(r,\phi)$ must of course satisfy the Helmholtz equation, a feature automatically
     underwritten when $v(r,\phi\mp\phi_{0})$ individually do so.  This latter aspect follows at once
     by displacing $\zeta$ contour $\rule{0mm}{0mm}C_{1}\cup C_{2}\rule{0mm}{0mm}$ in (1)
     through the respective horizontal amounts $\phi\mp\phi_{0},$
     a gesture which exposes to view plane-wave structures
     $\exp(-ikr\cos\{\zeta-\phi\pm\phi_{0}\})$ in integrand numerators which, {\em{ipso facto}}, validate
     the Helmholtz constraint.
     More than this, in order for (2) to satisfy the $E/H$ boundary
     conditions, it suffices to require that $v(r,\psi)$ as a function of place holder variable $\psi$
     be symmetric around both $\psi=0$ and $\psi=\gamma.$
     \vspace{-2mm}


           This symmetry attribute is likewise easily verified by noting in Figure 2 that, apart from their sense of traversal,
     $C_{1}$ and $C_{2}$ can be freely taken as genuine images of one another under reflection
     of $\zeta$ across its origin, {\em{viz.,}} $\zeta\leftrightarrow -\zeta.$\footnote{From (1) the symmetry around $\psi=\gamma$ emerges as
     \begin{eqnarray}
       2\gamma v(\gamma-\phi)& = & \int_{\atop{\rule{-2mm}{5mm}C_{1}}}
                              \frac{e^{-ikr\cos\zeta}}{1+e^{-i\pi(\zeta-\phi)/\gamma}}\,d\zeta-
                       \int_{\atop{\rule{-2mm}{5mm}C_{1}}}
                              \frac{e^{-ikr\cos\zeta}}{1+e^{i\pi(\zeta+\phi)/\gamma}}\,d\zeta  \nonumber \\
                             & = & \int_{\atop{\rule{-2mm}{5mm}C_{1}}}
                                   e^{-ikr\cos\zeta}\frac{e^{i\pi(\zeta-\phi)/\gamma}}{1+e^{i\pi(\zeta-\phi)/\gamma}}\,d\zeta-
                                   \int_{\atop{\rule{-2mm}{5mm}C_{1}}}
                                   e^{-ikr\cos\zeta}\frac{e^{-i\pi(\zeta+\phi)/\gamma}}{1+e^{-i\pi(\zeta+\phi)/\gamma}}\,d\zeta  \nonumber \\
                             & = & \int_{\atop{\rule{-2mm}{5mm}C_{1}}}
                              \frac{e^{-ikr\cos\zeta}}{1+e^{-i\pi(\zeta+\phi)/\gamma}}\,d\zeta-
                       \int_{\atop{\rule{-2mm}{5mm}C_{1}}}
                              \frac{e^{-ikr\cos\zeta}}{1+e^{i\pi(\zeta-\phi)/\gamma}}\,d\zeta  \nonumber \\
                             & = & 2\gamma v(\gamma+\phi)    \nonumber
    \end{eqnarray}
    and similarly for the symmetry around $\psi=0$ and all other prestidigitations inducing contour interchanges
    $C_{1}\leftrightarrow C_{2},$ as needed, but without further elaboration, in Eqs. (4)-(5) below.}
     But then $v(r,-\phi-\phi_{0})=v(r,\phi+\phi_{0})$ and $v(r,-\phi+\phi_{0})=v(r,\phi-\phi_{0}),$
     and similarly around $\gamma,$ in consequence of which $u(r,\phi)$ from (2) with its upper, negative sign on the
     right is antisymmetric around $\phi=0$ and $\phi=\gamma,$ being thus null and hence adequate to
     assure a tangential electric boundary condition at both wedge faces.
     And, for identically the same reasons,
     a choice of plus sign on the right in (2) provides a combination symmetric around both
     $\phi=0$ and $\phi=\gamma,$ whose presumably nonvanishing values there are well adapted to a
     purely tangential magnetic boundary condition.     
     We observe finally that the confirmed symmetry of $v(r,\psi)$ around both $\psi=0$ and
     $\psi=\gamma$ automatically generates, through endless ratcheting that alternates around these
     two points, a periodicity in the amount $\psi=2\gamma,$ {\em{viz.,}}
     $v(r,\psi+2\gamma)=v(r,\psi),$ already evident from the makeup of (1).
     \vspace{-3mm}
     
        It is to Pauli [{\bf{7}}] that we owe the seemingly innocent but actually quite crucial
     observation regarding the symmetry of $v(r,\psi)$ around both $\psi=0$ and $\psi=\gamma.$
     To a certain extent this liberates
     the discussion from having to drag in allusions to nonphysical Riemann sheets, and to entertain
     the mathematical fiction of primary excitations departing from physical space across sheet
     boundaries while specular reflections materialize in the opposite sense.  Mesmerizing
     and widely popular such scenarios may well be, but they tend to envelop the diffraction phenomenon
     amid some sort of ethereal aura which obscures physical understanding.
     \vspace{-3mm} 
          
          Most of these solution properties are deftly encapsulated in [{\bf{8}}], while a more complete
     discussion can be traced from [{\bf{9}}].  In [{\bf{5}}] and [{\bf{9}}], and to some extent
     also in [{\bf{6}}], one is exposed to considerable noise
     \newpage
     \mbox{    }
     \newline
     \newline
     \newline
     regarding a Riemann surface associated
     with the $2\gamma$ periodicity.  In our opinion, the fixed, universal aspect of Figure 2,
     and a simple recognition of the two symmetries around $\psi=0$ and $\psi=\gamma,$ with a
     $2\gamma$ periodicity noted as an automatic, albeit quite incidental consequence, more than
     suffice, and that thrashing around the mathematical concept of a Riemann surface simply beclouds
     the intended physical context.
     \vspace{-4mm}

     \section{Surface current density on wedge faces}
     \vspace{-3mm}     
          We set $\tau=\pm$ respectively for the upper ($\phi=0$) or lower ($\phi=\gamma$) face and, on the
     strength of the Faraday/Amp\`{e}re equations combined, find a $z$-directed current
     \begin{equation}
     K^{(\tau)}_{z}(r)=\frac{i\tau}{\omega\mu r}\frac{\partial E_{z}}{\partial\phi}\,,
     \end{equation}
     wherein the presence of $\tau$ as a multiplier acknowledges that the external magnetic field, with just
     a single, radial component, loops in opposite senses above and below.  Symbol $\mu$ denotes the
     uniform magnetic permeability of the ambient space, and sub/superscripts have been permitted to
     blossom by way of geometric memory prompts.  From (1) and (2), from the obvious reflection symmetry when
     $\zeta\leftrightarrow-\zeta$ between contours $C_{1}$ and $C_{2}$ as already noted, and with strategic
     appeal now and then to contour sense reversal, we readily find that
     \begin{equation}
     K^{(\tau)}_{z}(r)=\frac{\pi}{\omega\gamma^{2}\mu r}\left\{\rule{0mm}{5mm}L^{(\tau)}_{z}(r,\phi_{0})-
                                                              L^{(\tau)}_{z}(r,-\phi_{0})\right\} \,,
     \end{equation}
     with         
     \begin{equation}
     L^{(\tau)}_{z}(r,\pm\phi_{0})=\int_{\atop_{\rule{-2mm}{2mm}C_{1}}}e^{-ikr\cos\zeta}
     \frac{e^{i\pi(\zeta\,\mp\,\phi_{0})/\gamma}}
     {\rule{0mm}{3mm}(1-\tau e^{i\pi(\zeta\,\mp\,\phi_{0})/\gamma})^{2}}\,d\zeta\,.
     \end{equation}
     \vspace{-8mm}            
     \section{Surface current decomposition into a series of Bessel functions}
     \vspace{-3mm}     
          On the contour $C_{1}$ we are entitled to seek for the integrand in (5) a convergent
     power series development, one that opens the door to much useful processing.  Easily gotten as the
     first derivative of a suitable geometric series, this latter reads
     \begin{equation}
     L^{(\tau)}_{z}(r,\pm\phi_{0})=\tau\sum_{n=1}^{\infty}n\tau^{n}\int_{\atop_{\rule{-2mm}{2mm}C_{1}}}e^{-ikr\cos\zeta}
     e^{in\pi(\zeta\,\mp\,\phi_{0})/\gamma}d\zeta\,,
     \end{equation}     
     and, because
     \begin{equation}
     J_{p}(kr)=-\frac{1}{2\pi}e^{ip\frac{\pi}{2}}
        \int_{\atop_{\rule{-2mm}{2mm}C_{1}}}e^{-i(kr\cos\zeta- p\, \zeta)}d\zeta\,,
     \end{equation}             
      with $J_{p}$ being a Bessel function whose index $p$ need not be integral [{\bf{10}}], it finally condenses
      into
      \begin{equation}
      L^{(\tau)}_{z}(r,\pm\phi_{0})=-2\pi\tau\sum_{n=1}^{\infty}n\tau^{n}e^{-i\frac{n\pi}{\gamma}
      	\left(\frac{\pi}{2}\pm\phi_{0}\right)}J_{\frac{n\pi}{\gamma}}(kr)\,.
      \end{equation}
      One may remark in passing that, in seeking a series development at this point, we are in a sense unraveling,
      but only momentarily so, the path to solution adopted in [{\bf{4}}].
      \newpage
      \mbox{  }
      \vspace{-4mm}    
      \section{Sheet radiation determined by the Fourier transform of its current source}
      \vspace{-4mm}      
           Once currents $K^{(\tau)}_{z}(r),$ $\tau=\pm,$ are duly in hand,\footnote{We stress yet again
     that these currents have arisen in response to the {\em{total}} field, incident superposed upon that
     self-consistently radiated in the context of the {\em{bona fide}}, metallic wedge.} we proceed with
     field computation on the premise that henceforth they radiate, both up and down, within the ambient
     medium characterized by magnetic permeability $\mu$ as already introduced, dielectric permittivity
     $\epsilon,$ and a vanishing conductivity $\sigma\!\downarrow\! 0+,$ all three parameters being deemed
     to be uniform in both space and time.  We introduce moreover generic Cartesian co\"{o}rdinates
     $\{\hat{x},\hat{y}\},$ tailored as necessary to the two laminae:
     \begin{equation}
     \left(\!\begin{array}{c}
              \hat{x}\\
              \hat{y}
           \end{array}\!\right)=\left(\!\begin{array}{c}
                                   r\cos\phi \\
                                   r\sin\phi
                                    \end{array}\!\right)         
     \end{equation}  
     when $\tau=+,$ and
     \begin{equation}
     \left(\!\begin{array}{c}
     \hat{x}\\
     \hat{y}
     \end{array}\!\right)=\left(\!\begin{array}{c}
     r\cos(\phi-\gamma) \\
     r\sin(\phi-\gamma)
     \end{array}\!\right)         
     \end{equation}
     when instead $\tau=-.$  With this notation, which tethers all geometry to the master
     $\{r,\phi\}$ polar co\"{o}rdinates implied in Figure 1, we can resolve both radiated fields in terms
     of their respective spectra $A^{(\tau)}(\eta)$ of generalized plane waves
     \begin{equation}
     E^{(\tau)}_{z}(\hat{x},\hat{y})=\int_{-\infty}^{\infty}e^{i\eta\hat{x}-|\hat{y}|\sqrt{\eta^{2}-k^{2}\,}}
     	                              A^{(\tau)}(\eta)\,d\eta
     \end{equation} 
     in both their propagating (when $\eta^{2}<k^{2})$ and evanescent (when $\eta^{2}>k^{2}$) incarnations.
     Strictly speaking
     \begin{equation}
     k=\lim_{\atop{\sigma\downarrow 0+}}\omega\sqrt{\epsilon\mu\{1+i\sigma/(\omega\epsilon)\}\,}
     \end{equation}
     has a vanishing imaginary part so that the propagating/evanescent transition makes sense only when
     understood, by common convention, as the corresponding real limit.  Furthermore, when, as here, angular
     frequency $\omega$ is taken as positive, branch cuts for the square root may be taken up and down
     respectively from $\pm k,$ optimally approaching the imaginary axis.\footnote{The preferred branch cuts
     join Riemann sheets on which $\Re \sqrt{\eta^{2}-k^{2}\,}$ is uniformly of opposite sign.  The cuts
     themselves are thus defined by $\Re \sqrt{\eta^{2}-k^{2}\,} = 0.$  This has the pleasant consequence
     that contours may be dilated into upper/lower semicircular arcs at infinity which, of themselves, beget
     no contributions, the end results, save for residues, if any, being the residual loops around said
     branch cuts.  All of this, however, is without bearing upon our further arguments, as a result of which
     we have simplified our Figures 3-5 below by portraying branch cuts that point straight up or down.
     The {\em{raison d'\^{e}tre}} for such branch cut geometry is more full explicated in [{\bf{11}}], pp. 20-23.}

           Now, since currents $K^{(\tau)}_{z}(r),$ $\tau=\pm,$ are here viewed as disembodied entities
     floating in the ambient space, we can harness the Faraday-Amp\`{e}re duo along either sheet with
     $\hat{y}=0$ in the form
     \begin{equation}
     \frac{2}{i\omega\mu}\int_{-\infty}^{\infty}e^{i\hat{x}}\sqrt{\eta^{2}-k^{2}\,}A^{(\tau)}(\eta)\,d\eta=
     \left\{\rule{0mm}{7mm}\begin{array}{ccc}
                      0   &  ; &   \hat{x} < 0  \\
                      K^{(\tau)}_{z}(\hat{x}) &  ;  &   \hat{x} > 0
            \end{array}   \right.      
     \end{equation}
     requiring no further qualification as to magnetic field direction and yielding
     \begin{equation}
     A^{(\tau)}(\eta)=\frac{i\omega\mu}{4\pi\sqrt{\eta^{2}-k^{2}\,}}
              \int_{0}^{\infty}e^{-i\eta r} K^{(\tau)}_{z}(r)\,dr   
     \end{equation}
     under inverse Fourier transformation.              
     \newpage
     \mbox{     }          
         
     \section{Invoking the Weber-Schafheitlin integral}
          
          Reference to Eqs. (4) and (8) shows that (14) obliges us to consider next the Weber-Schafheitlin
     integral
     \begin{equation}
     M_{n}=\int_{0}^{\infty}e^{-i\eta r}r^{-1}J_{\frac{n\pi}{\gamma}}(kr)\,dr\,.
     \end{equation}      
     An adaptation of the work in [{\bf{12}}] then readily gives, with $\eta$ temporarily regarded as purely real,
     \begin{equation}
     M_{n}=\left(\frac{\gamma}{n\pi}\right)k^{-\frac{n\pi}{\gamma}}\times\left\{\rule{0mm}{10mm}\begin{array}{rcl}
          e^{-i\frac{n\pi^{2}}{2\gamma}}\left\{\rule{0mm}{5mm}\eta-\sqrt{\eta^{2}-k^{2}\,}\right\}^{\frac{n\pi}{\gamma}}
                                                          &   ;  &  |\eta|\leq k   \\
     e^{-i\frac{n\eta\pi^{2}}{2|\eta|\gamma}}
               \left\{\rule{0mm}{5mm}|\eta|-\sqrt{\eta^{2}-k^{2}\,}\right\}^{\frac{n\pi}{\gamma}}
                                                          &   ;  &  |\eta| > k    \; \,.
                                                                             \end{array} \right.  
     \end{equation}
     \vspace{-4mm}
     
     Anticipating next the summation stipulated by (8), we see that the overt index multiplier $n$
     conveniently disappears, and that convergence is quite unhindered for $|\eta|>k.$  This latter
     assertion follows by setting $|\eta|=k\cosh\theta,$ real $\theta>0,$ whence also
     \begin{eqnarray}
     |\eta|-\sqrt{\eta^{2}-k^{2}\,}  & =  & k\left\{\rule{0mm}{4mm}\cosh\theta-\sinh\theta\right\} \nonumber  \\
                                     & =  & ke^{-\theta}\,.
     \end{eqnarray}      
     For $|\eta|\leq k,$
     by contrast, since a similar parametrization as $\eta=k\cos\vartheta$ over $0\leq\vartheta\leq\pi$ yields
     \begin{equation}
     \eta-\sqrt{\eta^{2}-k^{2}\,}=ke^{i\vartheta} \,,
     \end{equation}
     it follows that convergence is assured by permitting $\vartheta$ to migrate {\em{upward}} in its own
     complex plane, $\vartheta=\alpha+i\beta,$ $0\leq\alpha\leq\pi,$ $\beta>0,$ an excursion which, with
     a view to     
     \begin{eqnarray}
     \eta &  =  &  k\cos(\alpha+i\beta)  \nonumber \\
          &  =  &  k\left\{\rule{0mm}{4mm}\cos\alpha\cosh\beta-i\sin\alpha\sinh\beta\right\} \;,
     \end{eqnarray}
     has the effect of depressing the Fourier spectral contour $\Gamma$ for Eq. (11)
     {\em{downward}} when $-k < \Re\,\eta < k .$\footnote{A {\em{downward}} contour shift of this sort is
     compatible with the analyticity prospects suggested in both (14) and (19).}
     The resulting contour, complete with its downward bulge, appears in Figure 3
     as a composite of three contiguous pieces, $\Gamma=\Gamma_{<}\cup\Gamma_{0}\cup\Gamma_{+}.$  Figure 3
     serves in addition to presage the existence of a simple pole ({\bf{P}}, colored green and situated to the left of the
     imaginary axis as befits the arbitrary choice $\phi_{0}<\pi/2$ depicted in Figure 1) and a saddle point
     ({\bf{SP}}, colored sandy beige), both lying on the real $\eta$ axis.  In particular, their relative order will shortly
     be seen to dictate two distinct categories of preferred deformations for $\Gamma,$ deformations which
     convey a most natural distinction between diffracted, and the dominant, reflected/shielding fields.

          Of course, once finite summations have been duly attained under the careful gaze of (18)-(19),
     we are given a {\em{carte blanche}} to exploit the benefits of contour deformation as far as analyticity
     may allow.  Such analyticity will indeed persist across regions where, strictly speaking, (18)-(19),
     when taken at face value, would predict outright catastrophes.

     \newpage
     \mbox{     }
          \vspace{-5mm}      
     \begin{center}
     	\includegraphics[angle=0.52,width=0.5\linewidth]{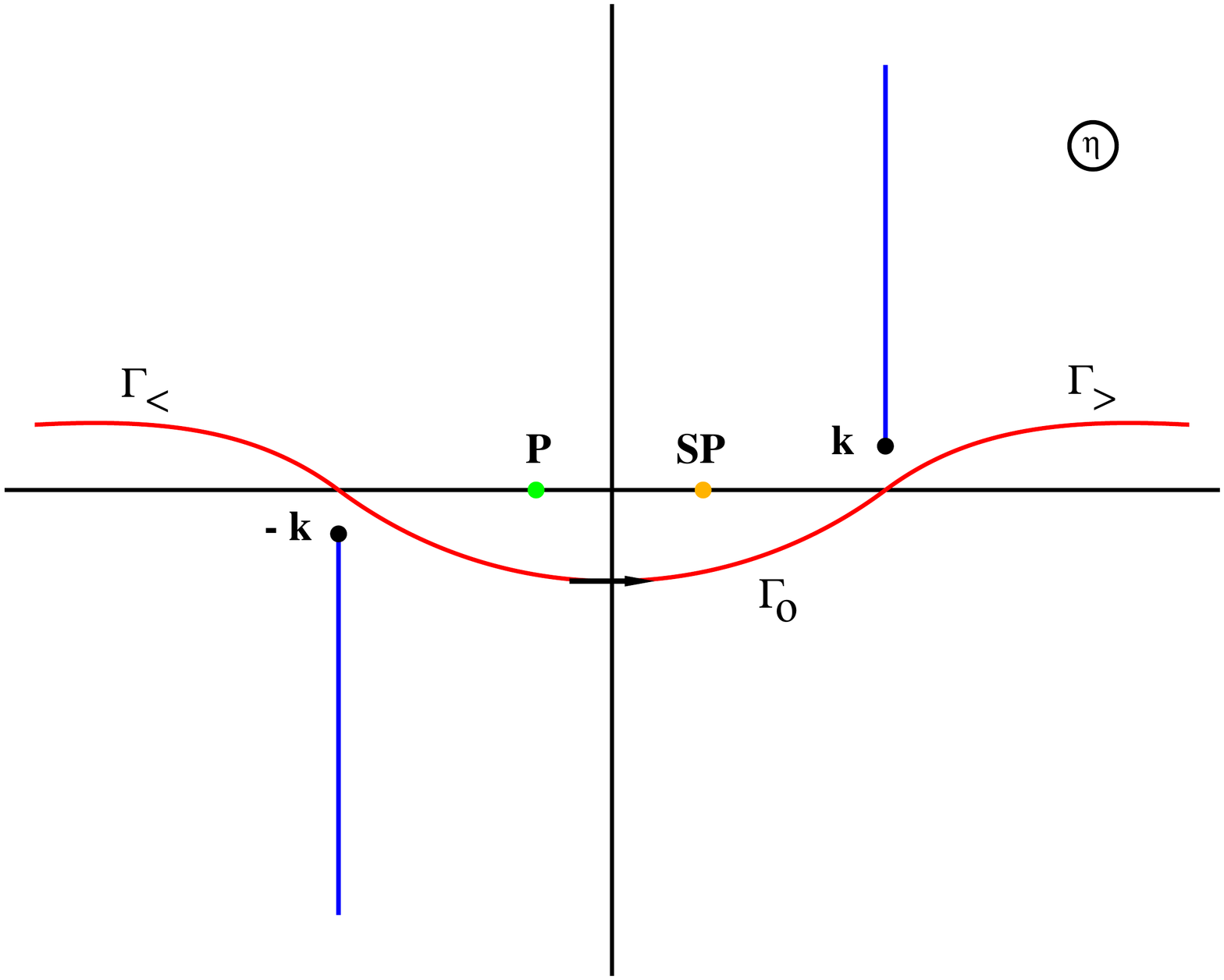}
     \end{center}                   
     \vspace{-12mm}
     \begin{center}
     	Figure 3.  Fourier spectral contour $\Gamma=\Gamma_{<}\cup\Gamma_{0}\cup\Gamma_{+}$ for Eq. (11).
     \end{center}

     \section{Series summation in reverse}
     
           Now that the radial quadrature mandated in (14) has been disposed of, our next step in the
     buildup of spectral amplitudes $A^{(\tau)}(\eta)$ is to perform the sums
     \begin{equation}
     N^{(\tau)}_{\pm}(\eta)=-2\pi\tau\left(\frac{\pi}{\omega\gamma^{2}\mu}\right)\sum_{n=1}^{\infty}
     n\tau^{n}e^{-i\frac{n\pi}{\gamma}\left(\frac{\pi}{2}\pm\phi_{0}\right)}M_{n}
     \end{equation}
     on which the interplay of (4) and (8) insists.  A glance at (16) reveals that these are again in the
     nature of geometric series, which give
     \begin{equation}
     N^{(\tau)}_{\pm}(\eta)=-\left(\frac{2\pi}{\omega\gamma\mu}\right)
     e^{-i\frac{\pi}{\gamma}(\pi\pm\phi_{0})}k^{-\frac{\pi}{\gamma}}\times\left\{\rule{0mm}{1.8cm}
     \begin{array}{lcl}
     \frac{\left\{\rule{0mm}{3mm}\eta-\sqrt{\eta^{2}-k^{2}\,}\right\}^{\frac{\pi}{\gamma}}}
     	{1-\tau e^{-i\frac{\pi}{\gamma}(\pi\pm\phi_{0})}k^{-\frac{\pi}{\gamma}}
     		\left\{\rule{0mm}{3mm}\eta-\sqrt{\eta^{2}-k^{2}\,}\right\}^{\frac{\pi}{\gamma}}}  &  ;  &  |\eta| \leq k  \\
     	&   &    \\
     	\frac{e^{i\frac{\pi^{2}}{2\gamma}(1-\eta/|\eta|)}
     		\left\{\rule{0mm}{3mm}|\eta|-\sqrt{\eta^{2}-k^{2}\,}\right\}^{\frac{\pi}{\gamma}}}
     		{1-\tau e^{-i\frac{\pi}{\gamma}(\pi\pm\phi_{0})}e^{i\frac{\pi^{2}}{2\gamma}(1-\eta/|\eta|)}k^{-\frac{\pi}{\gamma}}
      	   \left\{\rule{0mm}{3mm}|\eta|-\sqrt{\eta^{2}-k^{2}\,}\right\}^{\frac{\pi}{\gamma}}}  &  ;  &  |\eta| >  k  \;\,.
    	\end{array}  \right.
      \end{equation}
      
      \section{Field recovery:  pole and saddle points}
      
            On collating Eqs. (4), (8), (11), (14), (16), and (20), in that order, we are led to consider the auxiliary
      quantities
      \begin{equation}
      O^{(\tau)}_{\pm}(\hat{x},\hat{y})=\frac{i\omega\mu}{4\pi}\int_{\atop{\Gamma}}e^{i\eta\hat{x}-|\hat{y}|\sqrt{\eta^{2}-k^{2}\,}}
      \frac{N^{(\tau)}_{\pm}(\eta)}{\sqrt{\eta^{2}-k^{2}\,}}\,d\eta
      \end{equation}

      \newpage
      \mbox{    }
      \mbox{    }
      \newline
      \newline
      \newline
      whose difference 
      \begin{equation}
      E^{(\tau)}_{z}(\hat{x},\hat{y})=O^{(\tau)}_{+}(\hat{x},\hat{y})-O^{(\tau)}_{-}(\hat{x},\hat{y})
      \end{equation}
      provides at length the field radiated by the upper/lower wedge face respectively as $\tau=\pm.$
      
          We shall not attempt any evaluation of (23) for observation points $\{\hat{x},\hat{y}\}$ close to
     the diffractive edge.  Nowadays such evaluation can, if all else fails, be reasonably undertaken through
     outright numerical quadrature.  Physical insight greater by far is bestowed on (23) by passing at once to
     sufficiently remote field points and recognizing that (21) and (22) taken together usher in a simple pole,
     fixed in place by incidence angle $\phi_{0},$ and a pair of saddle points which respond to movement on
     the part of observation angle $\phi.$  An approach to, and ultimate transit across that simple pole,
     on the part of one such saddle point, undertaken from both directions, has a vivid interpretation as a
     shadow boundary crossing.  Such transit occurs at both {\em{bona fide}}
     shadow ($II-III$:  $\phi_{geom\,opt}=\pi+\phi_{0}$)
     and specular reflection ($I-II$:  $\phi_{mirror}=\pi-\phi_{0}$) boundaries.

         Our computational strategy unfolds henceforth by deforming spectral contour $\Gamma$ so as to traverse
     each saddle point individually, their locations shortly to be characterized, along proper, steepest descent
     paths at a standard, $\pi/4$ declination.  A pole may or may not be crossed during this redeployment and,
     if it is, account is made of its residue.  This latter, residue part of the calculation is not inherently
     asymptotic and, indeed, it will be seen that the two separate pole crossings individually switch, first off,
     and then on, the traditional reflected and the now eagerly sought shadowing fields, in that order, as $\phi$
     sweeps out its full range from $0$ to $2\pi.$
     \vspace{-3mm}
     
     \subsection{Simple pole fixed location}
     \vspace{-3mm}
     
          In (17) we had shown, when $|\eta|>k$ and with some real positive $\theta>0,$ that
     \begin{equation}     
     \left\{\rule{0mm}{3mm}|\eta|-\sqrt{\eta^{2}-k^{2}\,}\right\}/k=e^{-\theta}<1   \,,
     \end{equation}
     which clearly denies $N^{(\tau)}_{\pm}(\eta)$ any possibility of exhibiting real
     axis poles outside the canonical interval $-k<\eta<k$.  For such real
     axis poles one must accordingly turn to the second line in (19), which posits that, if they occur at all,
     they must do so when
     \begin{equation}
     1-\tau e^{-i\frac{\pi}{\gamma}(\pi\pm\phi_{0})}k^{-\frac{\pi}{\gamma}}
     \left\{\rule{0mm}{3mm}\eta-\sqrt{\eta^{2}-k^{2}\,}\right\}^{\frac{\pi}{\gamma}}=
     1-\tau e^{-i\frac{\pi}{\gamma}\left\{\rule{0mm}{2mm}\pi\pm\phi_{0}-\alpha^{(\tau)}_{p,\pm}\right\} }=0   \;.     
     \end{equation}
     For $\tau=+,$ the upper wedge face being designated as source, we clearly have $\alpha^{(+)}_{p,-}=\pi-\phi_{0}$
     as the single possible candidate.\footnote{The other candidate verifying $\pi+\phi_{0}-\alpha^{(+)}_{p,+}=0$
     or else $\alpha^{(+)}_{p,+}=\pi+\phi_{0}>\pi$ is disqualified by falling beyond the permissible interval.}
     For $\tau=-,$ by contrast, when it is the lower face which radiates, we require that
     \begin{equation}
     (\pi/\gamma)\times(\pi\pm\phi_{0}-\alpha^{(-)}_{p,\pm})=\pi     
     \end{equation}
     whence
     \begin{equation}
     \alpha^{(-)}_{p,\pm}=\pi\pm\phi_{0}-\gamma<0\,,
     \end{equation}
     both of which reside exterior to our preferred interval and thus cannot beget residues.

    \newpage
    \mbox{   }
    \mbox{   } 
    \newline
    
          We shall defer disclosing the residue results at simple pole
   $k\cos\left\{\alpha^{(+)}_{p,-}\right\}=k\cos(\pi\pm\phi_{0})=-k\cos\phi_{0}$,
    doubtless leaving the anxious reader, breathlessly riveted in temporary suspense, until the accompanying
    saddle point movement has been brought into sharper focus.
    \vspace{-3mm}
    
    \subsection{Saddle point moveable locations}
    \vspace{-3mm}
    
          Unlike the pole at $k\cos\left\{\alpha^{(+)}_{p,-}\right\},$ which is fixed once incidence direction
     $\phi_{0}$ has been specified, the saddle points $k\cos\left\{\alpha^{(\tau)}_{sp}\right\},$ while remaining
     indifferent to the $\pm$ partition called for in (23), do respond to both observation angle $\phi$ and wedge
     face marker $\tau.$ In contrast, the saddle point field contributions themselves respond to all available parameters.
     
         When field radius $r$ is sufficiently large, and together with it $\hat{x}$ and/or $\hat{y},$ or both, are
     similarly so in magnitude, one can resort to a saddle point approximation to the four quantities
     $O^{(\tau)}_{\pm}(\hat{x},\hat{y})$ subsumed beneath (22).  The two saddle points which now emerge are
     routinely found by requiring that the derivative of the phase in (22) vanish.  Thus 
     \begin{equation}
     \frac{d\left\{i\eta\hat{x}-|\hat{y}|\sqrt{\eta^{2}-k^{2}\,}\right\}}{d\,\eta}=0      
     \end{equation}
     whence it follows that
     \begin{equation}
     \hat{x}=\frac{\eta|\hat{y}|}{\sqrt{k^{2}-\eta^{2}\,}}\,.
     \end{equation}
     Reference to (9) and (10) then shows that (29) amounts to
     \begin{equation}
     \cos\alpha|\sin\phi|=\sin\alpha\cos\phi
     \end{equation}
     when $\tau=+,$ and
     \begin{equation}
     \cos\alpha|\sin(\phi-\gamma)|=\sin\alpha\cos(\phi-\gamma)
     \end{equation}
     if instead $\tau=-.$  And so it becomes a consistent gesture to set
     \begin{equation}
     \alpha^{(+)}_{sp}=\left\{\begin{array}{lcl}
     \phi       &  ;  &   0<\phi<\pi    \\
     2\pi-\phi  &  ;  &  \pi<\phi<2\pi  \,,
     \end{array}  \right. 
     \end{equation}
     and
     \begin{equation}
     \alpha^{(-)}_{sp}=\left\{\begin{array}{lcl}
     \phi-\gamma    &  ;  &  0<\phi-\gamma<\pi  \\
     2\pi-\phi+\gamma    &  ;  &  \pi<\phi-\gamma<2\pi  \,
     \end{array}  \right.
     \end{equation}
     by way of assuring that $0<\alpha^{(\pm)}_{sp}<\pi$ and hence that $\sin\alpha^{(\pm)}_{sp}>0.$
     Assured in fact are the more precise equalities $\sin\alpha^{(+)}_{sp}=|\sin\phi|$ and
     $\sin\alpha^{(-)}_{sp}=|\sin(\phi-\gamma)|$.

              We note that both saddle point locations $k\cos\left\{\alpha^{(\pm)}_{sp}\right\}$ traverse
      the canonical slot $(-k,k)$ twice, first backward and then forward,
      during the course of a full, $0<\phi<2\pi$ angular ambit.  During such full ambit, saddle point
      $k\cos\left\{\alpha^{(+)}_{sp}\right\}$ must contend with an encroachment, both fore and aft, upon simple
      pole $k\cos\left\{\alpha^{(+)}_{p,-}\right\},$ whereas $k\cos\left\{\alpha^{(-)}_{sp}\right\}$ is exempt
      from any such complication, there being now no corresponding
      \newpage
      \mbox{    }
      \newline
      \newline
      \newline
      poles $k\cos\left\{\alpha^{(-)}_{p,\pm}\right\}$
      in play.  This distinction finds a graphic embodiment in the fact that steepest descent passage through
      $k\cos\left\{\alpha^{(+)}_{sp}\right\}$ requires contour $\Gamma$ deformation of both categories depicted
      in Figures 4 and 5, whereas passage through $k\cos\left\{\alpha^{(-)}_{sp}\right\}$ follows Figure 5 only
      {\em{ad libitum}}, but of course with no pole (green dot) present.  
      For imminent use in Eqs. (41)-(44), and
      at considerable risk of belaboring the point, we emphasize that (32) and (33) can also be restated as
      \begin{equation}
      \left\{\begin{array}{lcl}
                \cos\alpha^{(+)}_{sp}=\cos\phi   &  ;  &  0 < \phi < 2\pi    \\
                \sin\alpha^{(+)}_{sp}=|\sin\phi| &  ;  &  0 < \phi < 2\pi
             \end{array}  \right.    
      \end{equation}
       and
      \begin{equation}
      \left\{\begin{array}{lcl}
      \cos\alpha^{(-)}_{sp}=\cos(\phi-\gamma)   &  ;  &  0 < \phi < 2\pi    \\
      \sin\alpha^{(-)}_{sp}=|\sin(\phi-\gamma)| &  ;  &  0 < \phi < 2\pi \;.
      \end{array}  \right.    
      \end{equation}
      \vspace{-6mm}
     \subsection{Saddle point pole crossings}
     \vspace{-2mm}     
           We can now summarize as follows the co\"{o}perative pole/saddle point evaluations as we traverse the full
     angular range, $0<\phi<2\pi,$ including in our mind's eye the wedge interior.  As we move outward from $\phi=0$
     in region $I$, the saddle point $\eta^{(+)}_{sp}=k\cos\left\{\alpha^{(+)}_{sp}\right\}$ slides steadily downward from $\eta^{(+)}_{sp}=k.$
     Saddle point crossing on behalf of $-O^{(+)}_{-}(r\cos\phi,r\sin\phi)$ in region $I$ can evidently occur only if
     inversion contour $\Gamma$ from Figure 3 had first been subjected to a deformation as conveyed in Figure 4,
     engendering an isolated loop around the pole at $\eta^{(+)}_{p,-}=k\cos\left\{\alpha^{(+)}_{p,-}\right\}.$
     The residue contribution
     from that pole is thus present throughout region $I$, wherein it conveys the specular, mirror companion of the incident
     plane wave, the latter only implicitly present in our radiative field buildup.  The co\"{e}xisting saddle point
     contribution is to be regarded hence as a diffractive, even if only an approximate correction.
     \vspace{-2mm}  
     
         As is universally recognized, the validity of this diffractive correction clearly deteriorates on saddle point
     approach to the pole, but, following observation point transit from region $I$ into region $II$ (regions $II$ and $III,$ and
     the wedge interior $\gamma<\phi<2\pi$ being deemed in shadow {\em{vis-\`{a}-vis}} the mirror field), regains its
     assigned level of validity.\footnote{People have devised various stratagems to mitigate saddle point failure in pole
     proximity.  One such can found in [{\textbf{13}}] and [{\textbf{14}}].  On the other hand, there is clearly no
     {\em{a priori}} physical reason for the apparent saddle point failure, one that entails a spurious divergence to
     infinity, and is simply an unintended artifact of pretending that a pole may be permitted to sit astride a steepest
     descent contour.  In fact, inversion contour $\Gamma$ from Figure 3 need not submit to the additional distortions
     in Figures 4 and 5, and does so only in the hope of procuring thereby some numerically useful and physically satisfying approximations.}
     \vspace{-2mm}
         
         Once $\phi$ enters region $II,$ the mirror/residue field is abruptly switched off, only to
     resume as a shielding field on entry into region $III.$  As in the $I\rightarrow II$ transition, localized caveats as
     to saddle point accuracy briefly flare into view around the moment of shielding/residue field ignition.  Indeed,
     the shielding field remains lit throughout region $III$ {\em{and}} the wedge interior combined,
     $\pi+\phi_{0}<\phi<2\pi.$  And then,
     once the incident field is at long last brought into play, it illuminates regions $I$ and $II$ and, across
     $\pi+\phi_{0}<\phi<2\pi$ (region $III$ plus wedge interior) it is extinguished by the shielding field.     
     \vspace{-2mm}

          Additional diffractive, saddle point radiative field contributions flow also from
     $O^{(+)}_{+}(r\cos\phi,r\sin\phi),$
     and $\pm O^{(-)}_{\pm}(r\cos\{\phi-\gamma\},r\sin\{\phi-\gamma\})$ across the full angular range, unburdened by any
     concern about pole
     \newpage
     \mbox{    }
     \newline
     \newline
     \newline
     encounters.  But of course the cumulative
     effect of them all, including that of $-O^{(+)}_{-}(r\cos\phi,r\sin\phi)$ under present view, does vanish exactly within
     wedge interior $\gamma<\phi<2\pi,$ as one easily verifies when equipped with the saddle point evaluations soon to follow.
     Such diffractive cancellation on wedge interior is itself a most welcome manifestation of radiative shielding.
     \vspace{-4mm}
    
          The distinction between the required deformations of spectral contour $\Gamma,$ before and after the first shadow crossing,
     $I\rightarrow II,$
     is illustrated in Figures 4 and 5.  Contour structure 4 clearly reverts to that of 5 on the subsequent entry $II\rightarrow III$
     into the classical shadow.  It bears repeating, perhaps, that these two deformation categories pertain only to the steepest
     descent path through $k\cos\left\{\alpha^{(+)}_{sp}\right\},$ while for partner $k\cos\left\{\alpha^{(-)}_{sp}\right\}$
     Figure 5 will suffice, with its green dot suppressed. 

     \vspace{-8mm}      
     \begin{center}
     	\includegraphics[angle=0.52,width=0.5\linewidth]{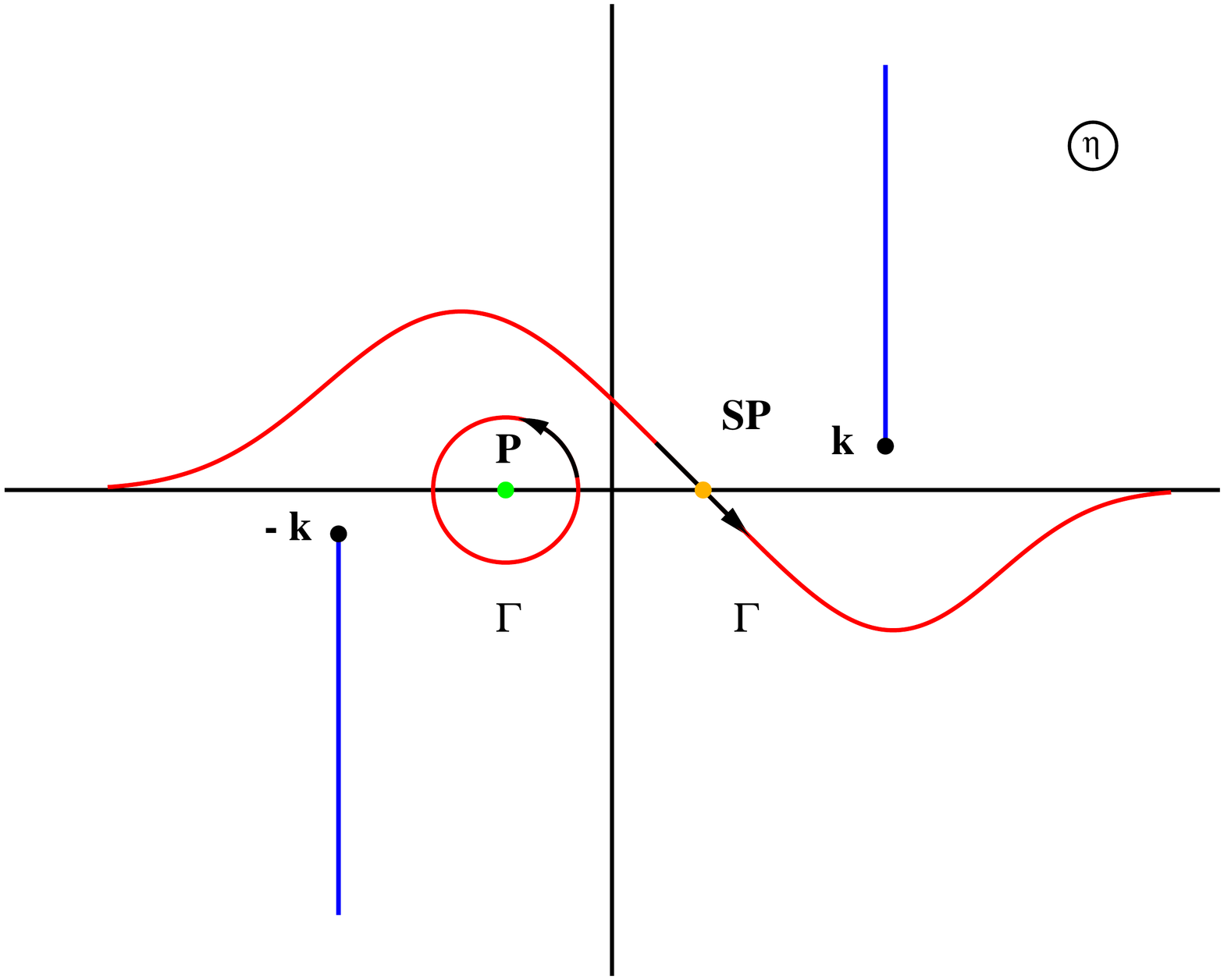}
     \end{center}                   
     \vspace{-11mm}
     \begin{center}
     	Figure 4.  Fourier spectral contour $\Gamma$ for Eq. (11) within mirror/shielding field\\regions $I$ and
     	 $III\,\cup\,\{\gamma<\phi<2\pi\}; \rule{3mm}{0mm}\tau=+$.
     \end{center}

     \vspace{-11mm}      
     \begin{center}
     	\includegraphics[angle=0.52,width=0.5\linewidth]{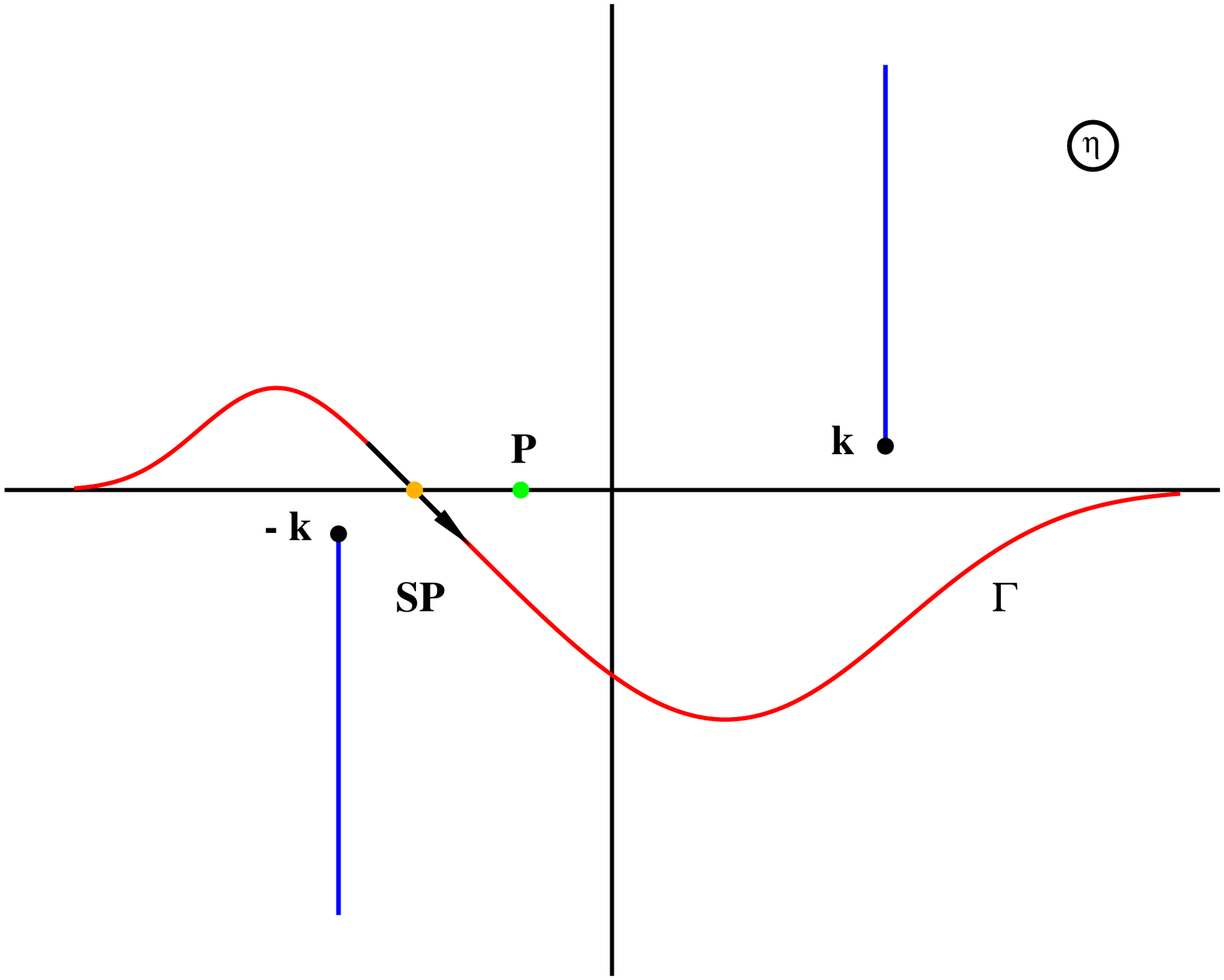}
     \end{center}                   
     \vspace{-11mm}
     \begin{center}
     	Figure 5.  Fourier spectral contour $\Gamma$ for Eq. (11) in directly illuminated\footnote{The
     qualifier {\em{``in directly illuminated region $II"$}} alludes to the traditional viewpoint, wherein region $II$ is indeed
     directly illuminated by the incident field, as of course so also is region $I,$ region $III\,\cup\,\{\gamma<\phi<2\pi\}$ in the
     meanwhile being traditionally consigned to a classic, geometric optics shadow, but, of course, with diffractive penetration into region
     $III$ still allowed.   That qualifying phrase should {\em{not}} in any way be
     construed as referring to field contributions radiated by the wedge surface currents under discussion.}
     region $II; \rule{3mm}{0mm}\tau=+.$
     \end{center}
     \newpage
     \mbox{    }
          
     \section{Shielding and mirror field contributions}
     
         And so, as we cross from region $II$ into either region $I$ or else region $III$ {\em{cum}} wedge, the pole
     contribution $E_{z,p}(r,\phi)$ lights up and, so to speak, remains lit.  Residue evaluation
     at $\eta^{+}_{p,-}=k\cos\left\{\alpha^{(+)}_{p,-}\right\}=k\cos(\pi-\phi_{0})=-k\cos\phi_{0}$ as guided by
     the first line from (21) yields, after an elementary albeit mildly tedious computation,
     \begin{equation}
     E_{z,p}(r,\phi)=-e^{-ikr(\cos\phi_{0}\cos\phi-\sin\phi_{0}|\sin\phi|)}\,,
     \end{equation}	
     which is nothing other than the specular, mirror field
     \begin{equation}
     E_{z,mirr}(r,\phi)=-e^{-ikr(\cos\phi_{0}\cos\phi-\sin\phi_{0}\sin\phi)}
                       =-e^{-ikr\cos(\phi+\phi_{0})}
     \end{equation}
     when $0<\phi<\pi-\phi_{0},$ and          
     \begin{equation}
     E_{z,shield}(r,\phi)=-e^{-ikr(\cos\phi_{0}\cos\phi+\sin\phi_{0}\sin\phi)}
                         =-e^{-ikr\cos(\phi-\phi_{0})}
     \end{equation}
     when instead $\pi+\phi_{0}<\phi<2\pi.$  A final, blanket superposition everywhere of the incident field
     \begin{equation}
     E_{z,inc}(r,\phi)=+e^{-ikr\cos(\phi-\phi_{0})}          
     \end{equation} 
     illuminates regions $I$ and $II$, and is fully extinguished by $E_{z,shield}(r,\phi)$ from (38) across region $III,$
     the traditional
     geometric optics shadow, and throughout the wedge interior, $\gamma<\phi<2\pi.$  Our mirror/shielding claims have
     thus been vindicated, at least at the dominant, plane wave level.
     
           It is somehow pleasing to one's physical sensibilities that both dominant fields, mirror and shielding,
     can be ascribed to just $-O^{(+)}_{-}(r\cos\phi,r\sin\phi),$ whose radiative source resides in a portion of the currents
     flowing across the upper, exposed wedge face.  That portion, moreover, is associated with the term $-v(r,\phi+\phi_{0})$
     present in (2) so as to maintain the angular symmetries that underlie the electric boundary condition now in force.  It
     is further associated with an image field emanating into real space from a nonphysical Riemann sheet.

     \section{Diffractive field contributions}
     
         An {\em{apropos}} moment has arrived to add in the diffractive, saddle point contributions
     $O^{(\tau,sp)}_{\pm}({\hat{x}},{\hat{y}})$ and to demonstrate that their sum
     \begin{equation}
     E^{(sp)}_{z}(r,\phi)=\sum_{\tau=\pm}\sum_{\zeta=\pm}\zeta\,O^{(\tau,sp)}_{z,\zeta}({\hat{x}},{\hat{y}})
     \end{equation}
     properly vanishes within wedge interior $\gamma<\phi<2\pi,$ whereas across the entire wedge exterior
     $0<\phi<\gamma,$ save for narrow angular exclusion slivers around shadow boundaries $\phi_{\pm,shad}=\pi\pm\phi_{0},$
     as previously described, it reproduces the standard foundation for the geometric theory of diffraction.
     \newpage
     \mbox{  }
     \mbox{  }
     \newline
             
        Adhering thus to a well worn recipe, we are instructed to approximate the phase $\Phi^{(\tau)}$ in (22) through the
      second order in the departure of $\eta$ from $\eta^{(\tau)}_{sp}=k\cos\left\{\alpha^{(\tau)}_{sp}\right\}.$
      For a phase $\Phi^{(\tau)}$ so curtailed we find
      \begin{equation}
      \Phi^{(\tau)}\approx ik{\hat{x}}\cos\left\{\alpha^{(\tau)}_{sp}\right\}
                     +ik|{\hat{y}}|\sin\left\{\alpha^{(\tau)}_{sp}\right\}-
                 \frac{i|{\hat{y}}|}{2k\sin^{3}\left\{\alpha^{(\tau)}_{sp}\right\}}
                 \left(\eta-k\cos\left\{\alpha^{(\tau)}_{sp}\right\}\right)^{2}
      \end{equation}
      which reveals a steepest descent direction aligned along $e^{-i\pi/4}$ and, by virtue of Eqs. (9)-(10) and
      (34)-(35) combined, simplifies to just
      \begin{equation}
      \Phi^{(\tau)}\approx ikr-\frac{i|{\hat{y}}|}{2k\sin^{3}\left\{\alpha^{(\tau)}_{sp}\right\}}
                                       \left(\eta-k\cos\left\{\alpha^{(\tau)}_{sp}\right\}\right)^{2}\,.
      \end{equation}
      The Gaussian integral which now confronts us is performed in routine fashion, the upshot of it all being that
      \begin{equation}
      O^{(\tau,sp)}_{z,\pm}({\hat{x}},{\hat{y}})=\frac{1}{\gamma}\sqrt{\frac{\pi}{2kr}\,}e^{i(kr-\pi/4)}
      \frac{e^{-i\pi(\pi\pm\phi_{0})/\gamma}\Lambda^{\pi/\gamma}_{\tau}}
      {\,1-\tau e^{-i\pi(\pi\pm\phi_{0})/\gamma}\Lambda^{\pi/\gamma}_{\tau}}
      \end{equation}
      with
      \begin{equation}
      \Lambda_{\tau}=\left\{\rule{0mm}{6mm}\begin{array}{lcl}
                          \cos\phi+i|\sin\phi|  &  ;  &   \tau = +   \\
                          \cos(\phi-\gamma)+i|\sin(\phi-\gamma)|  &  ;  &  \tau=-
                                    \end{array}   \right.
      \end{equation}
      itself discriminating between upper and lower wedge faces.
      \vspace{-2mm}
      
           The structure of (44) further forces us to distinguish between regions above and below with respect
      to each individual lamina.
       A complete, $0<\phi<2\pi$ circuit around the origin is thus partitioned in accordance with
       \vspace{-2mm}
     \begin{equation}
        \left\{\rule{0mm}{1.2cm}\begin{array}{lllcl}
         0<\phi<\gamma-\pi      & \longleftrightarrow & \Lambda_{+}=e^{i\phi}  & \&  & \Lambda_{-}=e^{i(\phi-\gamma+2\pi)}  \\
         \gamma-\pi<\phi<\pi    & \longleftrightarrow & \Lambda_{+}=e^{i\phi}  & \&  & \Lambda_{-}=e^{i(\gamma-\phi)}  \\
         \pi<\phi<\gamma        & \longleftrightarrow & \Lambda_{+}=e^{i(2\pi-\phi)} & \&  & \Lambda_{-}=e^{i(\gamma-\phi)}  \\
         \gamma<\phi<2\pi       & \longleftrightarrow & \Lambda_{+}=e^{i(2\pi-\phi)} & \&  & \Lambda_{-}=e^{i(\phi-\gamma)} \,.  
         \end{array}  \right.
        \end{equation}
  In addition to all previously announced phase requirements,
  there lurks implicitly in the background a need to situate all arguments upon the principal, $(-\pi,\pi)$ branch
  implicitly adopted throughout, so that exponentiation to the (generally) irrational power $\pi/\gamma$ may be consistently performed.
  The $2\pi$ phase shift in the first, third, and fourth lines stands in testimony to such adjustment.  
   Simple, albeit mildly tedious calculations based on (40), (43), and (45) inform us finally that
      \begin{equation}
      E^{(sp)}_{z}(r,\phi)=\sum_{\tau=\pm}\sum_{\zeta=\pm}\zeta\,O^{(\tau,sp)}_{z,\zeta}({\hat{x}},{\hat{y}})=0\,,
      \end{equation}
      a {\em{bona fide}} diffractive cancellation throughout the wedge interior, $\gamma<\phi<2\pi,$ whereas on its
      exterior, $0<\phi<\gamma,$ save for the exclusion slivers at shadow boundaries as previously mentioned, we find
 \begin{eqnarray}
      E^{(sp)}_{z}(r,\phi) & = & \sum_{\tau=\pm}\sum_{\zeta=\pm}\zeta\,O^{(\tau,sp)}_{z,\zeta}({\hat{x}},{\hat{y}})  \\
 & = & \frac{1}{\gamma}\sqrt{\frac{\pi}{2kr}\,}e^{i(kr+\pi/4)}\sin\left(\frac{\pi^{2}}{\gamma}\right)\times
   \left[\rule{0mm}{9mm}\frac{1}{\cos\left(\frac{\pi^{2}}{\gamma}\right)-\cos\left(\frac{\pi\{\phi-\phi_{0}\}}{\gamma}\right)}-
   \frac{1}{\cos\left(\frac{\pi^{2}}{\gamma}\right)-\cos\left(\frac{\pi\{\phi+\phi_{0}\}}{\gamma}\right)}\right] \,, \nonumber 
 \end{eqnarray}
 \vspace{-5mm}
 the very cradle indeed of GTD, with boundary conditions properly obeyed at both $\phi=0$ and $\phi=\gamma.$
 \newpage
 \mbox{  } 
 \vspace{-5mm}         
 \section{Magnetic field parallel to edge}
 \vspace{-4mm}
 
     When the incident, and therefore the entire magnetic field ${\textbf{H}}$ is purely edge directed, with but a
     single component $H_{z}$ (the so-called TM case, + sign in (2)), electric surface current ${\textbf{K}}$
shifts orientation from being purely axial to purely outgoing, perpendicular to the edge.  In the natural
co\"{o}rdinate system of either radiating face we confront now the surface current component $K_{\hat{x}}=
K_{r}$ and, as the analogue to spectral representation (11) we get
\vspace{-2mm}          
\begin{equation}
H^{(\tau)}_{z}(\hat{x},\hat{y})={\rm{sign}}(\hat{y})
\int_{-\infty}^{\infty}e^{i\eta\hat{x}-|\hat{y}|\sqrt{\eta^{2}-k^{2}\,}}C^{(\tau)}(\eta)\,d\eta \,,
\end{equation} 
with spectral amplitude $C^{(\tau)}(\eta)$ replacing the previous $A^{(\tau)}(\eta).$  A simple use of Amp\`{e}re's
law, followed by the obligatory Fourier inversion, permits one yet again to render spectral amplitude
$C^{(\tau)}(\eta)$ in terms of current density $K^{(\tau)}_{r}(r),$ {\em{viz.,}}
\begin{equation}
C^{(\tau)}(\eta)=\frac{1}{4\pi}\int_{0}^{\infty}e^{-i\eta r} K^{(\tau)}_{r}(r)\,dr   \,.
\end{equation}
Past this point the remainder of the program unfolds pretty much as before.  We recover the dominant shielding and
mirror field contributions {\em\`{a} la} Section 9, phrased now in terms of $H_{z}(r,\phi),$ the mirror field
confined to Region $I$ and the shielding extended across the augmented $III\,\cup\,$\{wedge interior\} angular
domain $\pi+\phi_{0}<\pi<2\pi.$  The saddle-point diffractive counterpart to (47)
\vspace{-2mm}
\begin{equation}
H^{(sp)}_{z}(r,\phi) = 
 \frac{1}{\gamma}\sqrt{\frac{\pi}{2kr}\,}e^{i(kr+\pi/4)}
\left[\rule{0mm}{9mm}
\frac{\sin(\pi^{2}/\gamma)}
{\cos\left(\frac{\pi^{2}}{\gamma}\right)-\cos\left(\frac{\pi\{\phi-\phi_{0}\}}{\gamma}\right)}+
\frac{\sin(\pi^{2}/\gamma)}
{\cos\left(\frac{\pi^{2}}{\gamma}\right)-\cos\left(\frac{\pi\{\phi+\phi_{0}\}}{\gamma}\right)}\right] 
\end{equation}
on wedge exterior $0<\phi<\gamma$ is gotten under a simple sign change from minus to plus on the right, and
co{\"{e}xists with a rigorously null magnetic field (at this level of approximation) on wedge interior
$\gamma<\phi<2\pi.$
\vspace{-11mm}
  
 \section{Parting comments}
\vspace{-5mm}
   Were we to relax the angular constraint, so that $\gamma/2>\phi_{0}>\gamma-\pi,$ then from (27) there would
appear a fresh simple pole with $\alpha^{(-)}_{p,+}=\pi+\phi_{0}-\gamma>0$ accompanied, on the physical side,
by additional mirror/shielding fields.  Such a scenario would require a similar, albeit more robust treatment
on its own terms, something that we have avoided in the interest of methodological simplicity.
\vspace{-2mm} 
 
     Furthermore, as we had previously mentioned, the present work, already at some modest level of intricacy, is but a
retrenchment, a reluctant retreat from the far more ambitious goal to which we had initially aspired, which is
to say, to actually base a wedge diffraction apparatus entirely upon an {\em{a priori}} null interior field
demand.  So magisterial a goal has so far proved to lie well beyond our reach.  And, if one is thus forced to
admit defeat when faced by a perfectly conducting obstacle, how much dimmer must be any solution prospects
when contemplating the permeable, dielectric wedge, one which no longer enjoys, even implicitly, anything
akin to the crucial symmetries\footnote{Despite our admittedly tepid enthusiasm for excessive preoccupation
with Riemann sheets, the symmetries at hand {\em{ipso facto}} compel one to acknowledge their presence
by nudging our gaze past wedge faces into the wedge interior.} $v(-\phi)=v(\phi)$ and
$v(\gamma-\phi)=v(\gamma+\phi)$?  Such perceived difficulties notwithstanding, we continue to entertain
the hope that exact solutions to scattering/diffraction by electromagnetically permeable, dielectric wedges
may yet be attained on the basis of field self-consistency, couched in the framework of integral equations
which superpose the self-consistent radiation from current sources distributed across wedge interiors.

 \newpage
 \mbox{  }      
 \section{Appendix:  a bibliographic potpourri}
 
     We wish to assemble here an informal, occasionally opinionated, {\em{ad hoc}} miniguide to the voluminous
     literature devoted to wedge diffraction and to its specialized half-plane subset gotten when $\gamma=2\pi.$
     Contemporary work, with much emphasis on wedge surface currents, is
     exemplified by [{\textbf{15}}].  Somewhat more recent is the review [{\textbf{16}}], valuable for both its
     richly illustrated content and an ample bibliography.  One of its three authors is Pyotr Yakovlevich Ufimtsev,
     the acknowledged father of the physical theory of diffraction, wherein physical emphasis is displaced, as here,
     from field boundary conditions as a prime focus of concern, to the actual surface currents that are the true
     sources of radiation [\textbf{17}].  At some remove in time from present activity are the book presentations
     [\textbf{18}] and [\textbf{19}], the latter, alas, in German, albeit published in Poland.  Reference [\textbf{20}]
     keeps alive the memory
     of Lamb's elegant solution for half-plane diffraction in parabolic co\"{o}rdinates.  His exquisitely
     concise analysis for the special case of perpendicular incidence upon a half plane is presented in [\textbf{21}].
     In [\textbf{22}] one finds a lively discussion of parabolic co\"{o}rdinates applied to diffraction, effectively
     illustrated and containing once more a robust bibliography to related work by Lamb, Credeli, Epstein, and even
     by Poincare.  Reference [\textbf{22}] is but one example in a tutorial archive embracing an absolutely
     phenomenal cornucopia of physical topics.
     
     Treatments of half-plane, $\gamma=2\pi$ diffraction via the Wiener-Hopf method run legion.  They are evolved
     along a traditional, integral-equation route in [\textbf{9}] ({\em{Chapter V.  Diffraction by a Plane Screen}}),
     then almost exclusively via the so-called	Jones method in [\textbf{11}] and [\textbf{23}].  The Jones method,
     in a nutshell, simply sidesteps from the very outset a cumbersome integral-equation intermediary by subjecting
     to Fourier transformation and boundary matching all underlying differential equations.  A fully explicit
     Wiener-Hopf diffractive solution in a traditional context, of considerable pedagogical value, is on offer
     in [\textbf{24}].
     
     An alternative to the Wiener-Hopf technique in the form of dual integral equations set amid a plane wave
     context has been evolved by P. C. Clemmow and is elaborated in [\textbf{14}].  A collateral presentation
     by him of similar material is found in the form of a collaborative contribution to [\textbf{25}].
     In [\textbf{26}] a brisk, Wiener-Hopf solution of the Sommerfeld diffraction problem provides a subordinate
     backdrop for comparison against Clemmow's dual integral equation method.  Reference [\textbf{27}] discusses
     at length sound pulse diffraction by a hard wedge, a substantially more intricate phenomenon than the pure
     time-harmonic scattering/diffraction entertained all along.
     
     A fresh impetus was imparted to wedge diffraction theory by the work of G. D. Maliuzhinets, who was able
     to generalize the Sommerfeld formalism so as to cope with surface boundary conditions more general than those
     of the present perfectly conducting default.  Moreover, the Maliuzhinets apparatus has some overlap with
     the Kontrorovich-Lebedev (K-L) transform, and with the work of T. B. A. Senior.  We are in no position to dwell
     on these developments, whose sources remain largely entombed in Russian language tomes and are thus
     difficult to acquire [\textbf{28}].  The anglophone literature is quite reticent in the matter of the K-L
     transform, the single available reference being [\textbf{29}], and that, too, in exceedingly scarce
     supply.  Some idea of the analytic complications encountered in applications of the
     Maliuzhinets/K-L program may be gleaned from [\textbf{30}], which likewise provides a valuable bibliography.
     It is fortunate indeed that explicit use of the K-L transform can be found in papers of high elegance,
     [\textbf{31}] and [\textbf{32}], which, while couched in \mbox{electrostatics, provide nevertheless a sound
     tutorial basis for application to the time-harmonic fields pres-}
          \newpage
          \mbox{   }
          \newline
          \newline
          \newline    
     ently of interest.\footnote{The qualifier
     {\em{Harmonic}} in the title of [\textbf{32}] alludes merely to Laplace's equation, and has no connection
     with time {\em{per se}}.}  The Maliuzhinets/K-L wedge analysis fervor remains undiminished.  Indeed, the
     1995 IEEE Antennas and Propagation Symposium devoted {\em{two}} 
     sessions to wedge scattering alone [\textbf{33}], and related work continues to appear in the electromagnetic
     literature.  For example, the explosive work of Vito Daniele
     and Guido Lombardi in Turin, Italy, among several others, advances at a furious pace, filling in every
     analytic nook and cranny of wedge theory, penetrable or not.  Reference [\textbf{34}], an adequate sample,
     sets the tone and high quality of these efforts.

     Still other references may be cited in support of the (null interior)/(self-consistent interior) field
     emphasis, references that provide elegant alternatives to contrast against the heavy labor incurred along
     traditional boundary matching lines.  Although these are not addressed to wedge diffraction {\em{per se}},
     they do underscore the benefits, both theoretical and numerical, that naturally accrue from basing one's
     arguments on field shielding/field self-consistency.  Thus, in [\textbf{35}] one finds such a solution
     to the problem of circular waveguide excitation by an azimuthally directed Dirac delta current source.
     Appeal to full shielding on the guide exterior yields the solution in a few deft steps which are simply
     breathtaking in their elegance and economy as compared to the algebraic avalanche of an earlier
     memorandum [\textbf{36}].  That latter tackled the analogous problem, but with a radially aligned point
     current source and a reliance on standard boundary conditions.  While ultimately adequate to its assigned
     task, it unleashed a torrent of algebra, a veritable mine field for potential mistakes, all of them
     studiously avoided.  Reference [\textbf{37}] assembles a small anthology of electromagnetic problems that are advantageously treated under a self-consistency viewpoint, whereas the power of
     electromagnetic self-consistency arguments even in a genuinely time-dependent setting finds confirmation
     in [\textbf{38}], wherein the problem of pulse impact upon a dielectric slab is easily disposed of once
     its time dependence has been subjected to Laplace transformation and the resulting framework recast in
     the guise of a self-consistent integral equation.


\section{References}
\parindent=0.0in
1.		J. Grzesik, {\bf{Field Matching through Volume Suppression}}, IEE Proceedings, Volume 127, Part H
(Antennas and Optics), Number 1, February 1980, pp. 20-26.

2.		J. A. Grzesik, {\bf{Wedge Diffraction as an Instance of Radiative Shielding}}, 1999 IEEE AP-S
International Symposium and USNC/URSI National Radio Science Meeting, July 11-16, Orlando, Florida.

3.		Arnold Sommerfeld, {\bfseries{Mathematische Theorie der Diffraction}}, Mathematische Annalen, Volume 16,
1896, pp. 317-374.

4.		H. M. MacDonald, {\bf{Electric Waves}}, Appendix D, Cambridge University Press, 1902, pp. 186-198.

5.		Arnold Sommerfeld in Philipp Frank and Richard von Mises (editors), {\bf{Die Differential- und
Integralgleichungen der Mechanik und Physik}}, {\em{Zweiter/physikalische/Teil}}, Friedrich Vieweg \& Sohn,
Braunchshweig, Deutschland, 1935, (Mary S. Rosenberg, WWII publisher, 235 West 108th Street, New York 25,
N. Y., 1943), {\bf{Theorie der Beugung}}, Chapter 20, pp. 808-830.
\newpage
\mbox{   }
\newline
\newline
\newline
6.		Arnold Sommerfeld, {\bf{Optics}}, Volume IV of {\em{Lectures on Theoretical Physics}}, translated from the
German by Otto Laporte and Peter A. Moldauer, Academic Press, New York, pp. 249-272.

7.		W. Pauli, {\bf{On Asymptotic Series for Functions in the Theory of Diffraction of Light}}, Physical Review,
Volume 54, December 1, 1938, pp. 924-931.

8.		L. D. Landau and E. M. Lifshitz, {\bf{Electrodynamics of Continuous Media}}, Volume 8 of {\em{Course of
Theoretical Physics}}, translated from the Russian by J. B. Sykes and J. S. Bell, Addison-Wesley Publishing
Company, Inc., Reading, Massachusetts, 1960, pp. 304-312.

9.		Bevan B. Baker and E. T. Copson, {\bf{The Mathematical Theory of Huygens Principle}}, Second Edition,
Oxford at the Clarendon Press, 1949, {\em{Chapter IV.  Sommerfeld's Theory of Diffraction}}, pp. 124-152;
{\em{Chapter V.  Diffraction by a Plane Screen}}, pp. 153-189.

10.		Julius Adams Stratton, {\bf{Electromagnetic Theory}}, McGraw-Hill Book Company, Inc., New York, 1941,
pp. 364-369.

11.		R. Mittra and S. W. Lee, {\bf{Analytical Techniques in the Theory of Guided Waves}}, The MacMillan
Company, New York, 1971, pp. 20-23 (preferred Riemann sheet), pp. 82ff, (Wiener-Hopf), pp. 97ff. (Jones method).

12.		G. N. Watson, {\bf{A Treatise on the Theory of Bessel Functions}}, Second Edition, Cambridge University
Press, 1966, p. 405, formulae (2) and (3).

13.		P. C. Clemmow, {\textbf{Some Extensions of the Method of Steepest Descents}},  The Quarterly Journal of
Mechanics and Applied Mathematics, Volume 3, Number 2, 1950, pp. 924-931.

14.		P. C. Clemmow, {\textbf{The Plane Wave Spectrum Representation of Electromagnetic Fields}},  The Institute
of Electrical and Electronics Engineers, Inc., New York, 1996, pp. 43-58, with particular attention to pp. 56-58.

15.		Adam Ciarkowski, Johannes Boersma, and Raj Mittra, {\textbf{Plane-Wave Diffraction by a Wedge--A Spectral
Domain Approach}}, IEEE Transactions on Antennas and Propagation, Volume AP-32, Number 1, 1 January 1984, pp. 20-29.

16.		Feray Haciveli\v{o}gliu, Levent Sevgi, and Pyotr Ya. Ufimtsev, {\textbf{Electromagnetic Wave Scattering from
a Wedge with Perfectly Reflecting Boundaries:  Analysis of Asymptotic Techniques}}, IEEE Antennas and Propagation
Magazine, Volume 53, Number 3, June 2011, pp. 232-253.

17.		Pyotr Ya. Ufimtsev, {\textbf{Fundamentals of the Physical Theory of Diffraction}}, John Wiley and Sons, Inc.,
Hoboken, New Jersey, 2007.

18.		Harry Bateman, {\textbf{Partial Differential Equations of Mathematical Physics}},
Cambridge University Press/Dover Publications, New York, 1959, {\em{Chapter XI.  Diffraction Problems}}, pp. 476-490.
\newpage
\mbox{   }
\newline
\newline
\newline
19.		A. Rubinowicz, {\textbf{Die Beugungswelle in der Kirchhoffschen Theorie der Beugung}}, Polska Akademia Nauk,
Monografie Fizyczne, Panstwowe Wydawnictwo Naukowe, Warszawa, 1957, {\em{IV.  Beugung an einer Halbebene}}, pp.
125-149.
\vspace{-1mm}

20.		H. Lamb, {\textbf{On Sommerfeld's Diffraction Problem, and on Reflection by a Parabolic Mirror}}, Proceedings
of the London Mathematical Society, Volume 4, 1906, p. 190ff.

21.		Sir Horace Lamb, {\textbf{Hydrodynamics}}, Sixth Edition, Dover Publications, New York, 1945, Section 308, pp. 538-541.
\vspace{-2mm}

22.		Kirk T. MacDonald, {\textbf{Sommerfeld's Diffraction Problem}}, Physics Examples, Princeton University, Department
of Physics, Joseph Henry Laboratories, Princeton, New Jersey, June 25, 2014, pp. 1-18.

23.		B. Noble, {\textbf{Methods based on the Wiener-Hopf Technique}}, Pergamon Press, New York, 1958.
\vspace{-1mm}

24.		Julian Schwinger, Lester L. DeRaad, Jr., Kimball A. Milton, and Wu-yang Tsai, {\textbf{Classical Electrodynamics}},
Perseus Books, Reading, Massachusetts, 1998, {\em{Section 48.  Exact Solution for Current}}, pp. 512-516.
\vspace{-2mm}

25.		Max Born and Emil Wolf, {\textbf{Principles of Optics}}, {\em{Electromagnetic theory of propagation, interference
and diffraction of light}}, Seventh (expanded) edition, Cambridge University Press, 2003, {\em{XI.  Rigorous
diffraction theory}}, pp. 633-672 (contributed by P. C. Clemmow).

26.		George F. Carrier, Max Krook, and Carl E. Pearson, {\textbf{Functions of a Complex Variable}}, {\em{Theory and
Technique}}, McGraw-Hill Book Company, New York, 1966, {\em{Section 8-5.  Dual Integral Equations}}, pp. 399-404.
\vspace{-2mm}

27.		F. G. Friedlander, {\textbf{{Sound Pulses}}, Cambridge Monographs on Mechanics and Applied Mathematics,
Cambridge University Press, 1958, {\em{}hapter 5.  The Diffraction of a Pulse by a Wedge}}, pp. 108-146.

28.		G. A. Grinberg, {\textbf{Selected Problems in the Mathematical Theory of Electric and Magnetic Phenomena}},
USSR Academy of Sciences, 1948, Chapter XXII.

29.		Ian N. Sneddon, {\textbf{The Use of Integral Transforms}}, McGraw-Hill Book Company, New York, 1972, {\em{Chapter 6,
The Kontorovich-Lebedev Transform}}, pp. 353-368.

30.		Ivan Joseph La Haie, {\textbf{Function-Theoretic Techniques for the Electromagnetic Scattering by a Resistive
Wedge}}, Technical Report, The University of Michigan, Department of Electrical Engineering, Radiation Laboratory,
Ann Arbor, Michigan 48109, September 1981, pp. 39-80.

31.		K. I. Nikoskinen and I. V. Lindell, {\textbf{Image Solution for Poisson's Equation in Wedge Geometry}},
IEEE Transactions on Antennas and Propagation, Voloume 43, Number 2, February 1995, pp. 179-187.

32.		Robert W. Scharstein, {\textbf{Green's Function for the Harmonic Potential of the Three-Dimensional
Wedge Transmission Problem}}, IEEE Transactions on Antennas and Propagation, Volume 52, Number 2, February 2004,
pp. 452-460.
\newpage
\mbox{   }
\newline
\newline
\newline
33.		IEEE Antennas and Propagation Society International Symposium, Newport Beach, California, June 18-June 23, 1995,
Symposium Digest, Volume Two, Session 6, Scattering by Wedges I, p. 1067ff., Session 9, Scattering by Wedges II, p. 1347ff.

34.		Vito G. Daniele and Guido Lombardi, {\textbf{Wiener-Hopf Solution for Impenetrable Wedges at Skew Incidence}},
IEEE Transactions on Antennas and Propagation, Volume 54, Number 9, September 2006, pp. 2472-2485.

35.		J. A. Grzesik, {\textbf{Vector Green's Function for the Circular Waveguide.  Part II:  Azimuthal Field Projection}},
Interoffice Correspondence, TRW Defense and Space Systems Group, One Space Park, Redondo Beach, California 90278,
2 September 1981, pp. 1-9.

36.		J. A. Grzesik, {\textbf{Vector Green's Function for the Circular Waveguide}},
Interoffice Correspondence (80) 4352-36, TRW Defense and Space Systems Group, One Space Park, Redondo Beach, California 90278,
18 September 1980, pp. 1-25.

37.		J. A. Grzesik and S. C. Lee, {\textbf{The Dielectric Half Space as a Test Bed for Transform Methods}}, Radio Science,
Volume 30, Number 4, July-August 1995, pp. 853-862.

38.		J. A. Grzesik, {\textbf{EM Pulse Transit across a Uniform Dielectric Slab}} 
(CEM-TD 2007, 7$^{{\rm{th}}}$ Workshop on Computational Electromagnetics in
Time Domain, October 15-17, 2007 - Perugia - Italy), Northrop Grumman Space Technology,
One Space Park, Redondo Beach, CA 90278, pp. 1-30.

\end{document}